\documentclass[a4paper,11pt]{article}
\usepackage{jheppub} 
\usepackage{lineno}

\usepackage{graphicx}
\usepackage{dcolumn}
\usepackage{bm}

\usepackage[T1]{fontenc}  
\usepackage[normalem]{ulem} 
\usepackage{graphicx}
\usepackage{tikz}       
\usepackage{subcaption}
\usepackage{diagbox}

\def\be{\begin{equation}}
\def\ee{\end{equation}}
\def\ba{\begin{eqnarray}}
\def\ea{\end{eqnarray}}

       \def\a {\alpha}  \def\d {\dot}              \def \inf {\infty}  
              \def\.{\cdot}

\title{Chaotic motion of particle in regular black hole supported by Galactic halo}

\author[1]{Aofei Sang\note{Corresponding author.}}
\affiliation{${}^1$Department of Physics, Southern University of Science and Technology, Shenzhen 518055, China}
\emailAdd{aofeisang@mail.bnu.edu.cn}
\emailAdd{12331027@mail.sustech.edu.cn}
\author{Ziyi Fang${}^{2}$}
\affiliation{${}^2$Department of Electrical and Electronic Engineering, Southern University of Science and Technology, Shenzhen 518055, China}

\abstract{We investigate the chaotic dynamics of the massless test particles moving in the regular black hole supported by a Dehnen-type dark matter halo. By limiting the particle within a external harmonic potential, we employ Poincaré sections and Lyapunov exponents as diagnostic tools and analyze the transition from regular to chaotic motion as the halo scale parameter $a$ increase. 
These findings indicate that the galactic halo acts as a primary driver of chaos in this regular black hole geometry, significantly distorting the phase space structure near the potential center, acting as a primary driver of chaos. 
Crucially, our results elucidate the distinct imprint of dark matter halos on particle dynamics, suggesting that observing such chaotic signatures in astrophysical systems could provide a novel method for detecting and constraining the properties of dark matter in future observations.

}

\begin{document}
\maketitle
\flushbottom

\section{Introduction}\label{sec1}

The problem of spacetime singularities remains one of the most profound conceptual challenges in general relativity. According to the Penrose-Hawking singularity theorems, under physically reasonable energy conditions and causal structure assumptions, gravitational collapse inevitably leads to curvature singularities, where geodesic incompleteness arises and curvature invariants diverge\cite{Penrose:1964wq,Hawking:1970zqf,Hawking:1973uf}. 
This signals the breakdown of the classical theory in the strong-field regime and indicates the necessity of either new physics or modified geometrical structures near the black-hole core. 

Constructing singularity-free(regular) black hole solutions is therefore not merely a mathematical exercise, but a fundamental step toward understanding the interplay between strong gravity and quantum effects.
The earliest regular black holes were proposed by Bardeen and Hayward, who introduced de-Sitter cores at the center to eliminate the curvature singularity while recovering the Schwarzschild solution at large distances\cite{Bambi:2013caa,Hayward:2005gi}. Subsequently, regular black holes supported by nonlinear electrodynamics\cite{Ayon-Beato:1998hmi,Bronnikov:2000vy,Bronnikov:2005gm}, anisotropic fluids\cite{Simpson:2018tsi,Fan:2016hvf}, and quantum gravity-inspired models\cite{Bambi:2013caa,Ashtekar:2018lag,Modesto:2008jz,Bonanno:2000ep} have been studied. These models aim to address the singularity problem by incorporating various theoretical mechanisms, such as the modification of the gravitational field at small scales due to quantum effects or the introduction of effective matter distributions. 

Another significant class of regular black holes is those caused by dark matter distributions\cite{Zhao:1995cp,Dehnen:1993uh,Navarro:1996gj,Kar:2025phe,Konoplya:2025ect,Dekel:2017bwy,Taylor:2002zd}. Recent studies, such as those by Konoplya and Zhidenko\cite{Konoplya:2025ect}, have demonstrated that dark matter halos can act as a natural mechanism for avoiding singularities, providing a physically motivated source for regular black hole solutions. 
In their construction, regularity is not imposed geometrically but emerges dynamically from the physical properties of the matter source\cite{Konoplya:2025ect}. Their analysis lies in identifying clear criteria under which halo-supported configurations admit event horizons while preserving asymptotic flatness. In other words, the avoidance of the singularity becomes a direct consequence of a finite central density in realistic dark matter halos. 
For comparison, the Simpson-Visser regular black hole~\cite{Simpson:2018tsi} represents a geometrically elegant regularization scheme obtained by replacing the radial coordinate according to $r \to \sqrt{r^2+a^2}$, thereby smoothing the Schwarzschild singularity. While this construction yields a simple analytic form and introduces a regular core controlled by the parameter $a$, the parameter does not directly correspond to a specific astrophysical density distribution.
Moreover, in Ref.\cite{Konoplya:2025ect,Lutfuoglu:2026fks}, the authors go beyond geometric construction by performing a stability analysis of axial perturbations and by computing observable quantities such as photon-sphere radii, shadow, and Lyapunov exponents. 
This development represents an important step in bridging theoretical models with realistic astrophysical environments, offering new possibilities for observational tests through black hole shadow measurements, gravitational wave signals and other related phenomena.



The study of particle motion near black holes is a crucial approach to understanding their properties. In particular, it has been observed that the presence of a black hole event horizon can lead to chaotic motion of particles, when these particles are initially moving in a harmonic potential. This phenomenon, first explored in the context of Schwarzschild black holes\cite{Dalui:2018qqv}, has been studied in various black hole spacetimes and gravity theories\cite{An:2025xmb,Cao:2024pdb,Cao:2025qpz,Dalui:2025bwm,He:2023dcz,Chen:2016tmr,Azreg-Ainou:2026xcc,Das:2025eiv,Das:2025vja}. 
Researchers have utilized techniques like Poincaré sections and Lyapunov exponents to quantify chaos and detect transitions from regular to chaotic motion in the near-horizon region. Recent work, such as that in Ref.\cite{Azreg-Ainou:2026xcc,Das:2025eiv,Das:2025vja}, has extended these investigations by considering the influence of dark matter halos on particle motion, finding that dark matter can significantly enhance the chaotic behavior. 
These studies reveals that the presence of a dark matter halo alters the effective potential around the black hole, thus influencing the stability of particle orbits and promoting chaotic dynamics as the particle approaches the event horizon. However, they mainly focus on singular black hole with Dehnen-($1,\,1/4,\,5/2$) type dark matter halo\cite{Das:2025eiv,Das:2025vja}. Furthermore, they explore how this chaotic behavior may manifest in extreme-mass-ratio inspirals(EMRIs), linking the onset of chaos to gravitational wave signals detectable by future space-based observatories like LISA. The results of these studies underline the critical role of dark matter in shaping the dynamics of black hole systems and suggest that chaotic particle motion could provide valuable insights into the observational characteristics of black holes.

In this paper, we will study the chaotic behavior of particles near the event horizon of regular black holes supported by dark matter halo of Dehnen-(1,4,0) type\cite{Konoplya:2025ect,Lutfuoglu:2026fks}. This research aims to enhance our understanding of the properties of regular black holes while highlighting the influence of dark matter on particle dynamics in their vicinity. Additionally, this study may provide insights into potential observational imprints in gravitational waves, which could offer further evidence of the effects of dark matter on black hole environments in the future.

This paper is organized as follows: In Sec.\ref{sec2}, we geometry of the regular black hole supported by a dark matter halo and gives the equation of motion for a massless particle moving in a harmonic potential. In Sec.\ref{sec3}, we solve the motion of the particle and show the poincar\'e section for some parameter. In Sec,\ref{sec4}, we discuss the Lyapunov exponent. Finally, we draw conclusion and give some discussion in sec.\ref{sec5}.

\section{regular black hole supported by a Galactic Halo}\label{sec2}



\begin{figure}[t]
    \centering
    \includegraphics[width=0.8\textwidth]{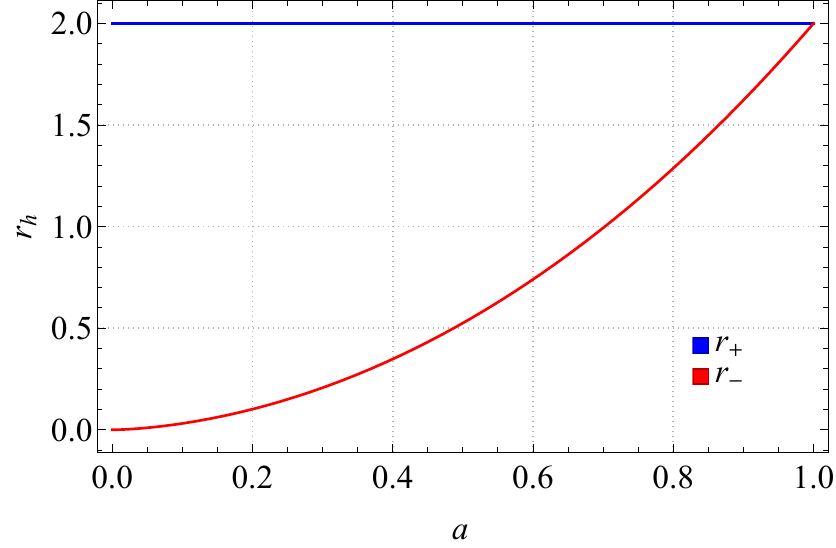}
    \caption{The inner($r_-$) and outer($r_+$) horizon of the regular black hole for $a\in (0,1]$}\label{fig-horizon}
  \end{figure}

In this paper, we consider a static, spherically symmetric spacetime sourced by an effective Dehnen-type dark matter halo described as an anisotropic fluid, which has been discovered that this dark matter field profile can lead to an exact regular black hole solutions.\cite{Konoplya:2025ect,Lutfuoglu:2026fks}.
Accordingly, we start from the standard Schwarzschild-like ansatz
\begin{equation}
ds^{2}=-f(r)\,dt^{2}+\frac{dr^{2}}{f(r)}+r^{2}\left(d\theta^{2}+\sin^{2}\theta\,d\phi^{2}\right),
\label{eq:sss_metric}
\end{equation}
where $f(r)$ is to be determined self-consistently from the Einstein field equations
\begin{equation}
G_{\mu\nu}=8\pi\,T_{\mu\nu}.
\label{eq:einstein}
\end{equation}

To model the dark matter distribution, we adopt an anisotropic-fluid energy-momentum tensor of the form
\begin{equation}
T^{\mu}{}_{\nu}=\mathrm{diag}\big(-\rho(r),\,P_{r}(r),\,P_{t}(r),\,P_{t}(r)\big),
\label{eq:stress_tensor}
\end{equation}
where $\rho$ is the energy density, and $P_r$ and $P_t$ are the radial and tangential pressures, respectively. 
With the metric \eqref{eq:sss_metric}, it is convenient to introduce a mass function $m(r)$ via
\begin{equation}
f(r)=1-\frac{2m(r)}{r}.
\label{eq:f_m}
\end{equation}
Then the $(t,t)$ component of \eqref{eq:einstein} yields the usual relation between $m(r)$ and the density profile,
\begin{equation}
m'(r)=4\pi r^{2}\rho(r),
\quad\Rightarrow\quad
m(r)=4\pi\int_{0}^{r}\rho(\tilde r)\,\tilde r^{2}\,d\tilde r,
\label{eq:mass_function}
\end{equation}
where regularity at the origin motivates $m(0)=0$.

We further impose the (radial) equation of state adopted in the paper,
\begin{equation}
P_r(r)=-\rho(r),
\label{eq:eos}
\end{equation}
which simplifies the remaining field equations and ensures that the core can approach an effective de Sitter behavior for suitable $\rho(r)$.
The tangential pressure $P_t$ is then determined by the angular components of \eqref{eq:einstein} together with stress-energy conservation, 
so that the system is closed once $\rho(r)$ is specified.

As the key physical input, we take the dark-matter halo density to follow a Dehnen-type profile\cite{Dehnen:1993uh,Das:2025eiv,Das:2025vja,Lutfuoglu:2026fks},
\begin{equation}
\rho(r)=\rho_0\left(\frac{r}{a}\right)^{-\gamma}\left(1+\frac{r^\a}{a^\a}\right)^{(\gamma-\beta)/\a}
\label{eq:dehnen}
\end{equation}
where $a>0$ is the characteristic halo scale radius, $\beta$ controls the inner slope, and $\rho_0$ sets the overall normalization. 
In this paper, we would like to adopt $\alpha=1$, $\beta=4$ and $\gamma=0$ to achieve a regular black hole.
Substituting \eqref{eq:dehnen} into \eqref{eq:mass_function} gives an analytic $m(r)$, and hence an analytic metric function through \eqref{eq:f_m}\cite{Konoplya:2025ect,Lutfuoglu:2026fks}
\begin{equation}
f(r)=1-\frac{2Mr^{2}}{(r+a)^{3}}\,.
\label{eq:model1}
\end{equation}
Here, $M$ is the ADM mass. The spacetime is asymptotically Schwarzschild ($f\to 1-2M/r$) when $r\to\infty$, while remaining regular at $r=0$ due to the softened mass profile $m(r)\sim r^{3}$ near the origin.
More generally, different Dehnen parameter choices lead to alternative expressions for $m(r)$ and thus for $f(r)$, but the derivation strategy remains the same. 

Then, we would like to interpret the structure of the regular black hole. Since we concentrate on a black hole, we assume there is a horizon, which is given by $f(r_1)=0$. 
Then, we can express the ADM mass in terms of the holo scale $a$ and the horizon:
\be\begin{aligned}
    M=\frac{(r_1+a)^3}{2 r_1^2}\,.
\end{aligned}\ee
Further, we the metric function can be cast to
\be\begin{aligned}
    f(r)=1-\frac{r^2}{r_1^2} \frac{(r_1+a)^3}{(r+a)^3}\,.
\end{aligned}\ee
Clearly, $r=r_h$ is one of the solution for $f(r)=0$. Without loss of generality, we would like to fix $r_1$ and set $r_1=2$. Then, it is easy to get the other positive real root of $f(r)=0$:
\be\begin{aligned}
   r_2= \frac{1}{8} \left(a^3+6 a^2+\sqrt{32 a^3+\left(a^3+6 a^2\right)^2}\right)
\end{aligned}\ee
When $a=0$, the spacetime reduce to Schwarzschild spacetime, which has one horizon $r_1=r_2=2$. When $a\in(0,1)$, $r_1>r_2$. 
We would like to denote the larger positive root $r_1$ as $r_+$, i.e. the outer horizon, and denote the other as $r_-$, which is the inner horizon. Moreover, it is easy to determine that the physical region with $f(r)>0$ is given by $r>r_+$. When $a=1$, the inner and outer horizons coincide, and the spacetime describes an extreme black hole. When $a>1$, $r_2$ becomes the outer horizon. In this paper, we would like to focus on $a\in[0,1]$, which is sufficient to illustrate the influence of dark matter halo on the chaotic of particle motion.
In Fig.\ref{fig-horizon}, we show how the horizon changes with the value of $a$.

This metric is introduced as a particularly simple analytic example of a regular black hole spacetime supported by a realistic galactic-halo matter profile. The construction is motivated by the idea that empirically used halo density distributions, supplemented with an appropriate pressure prescription, can yield exact asymptotically flat solutions in Einstein gravity that avoid curvature singularities, providing a more physically grounded alternative to regular metrics. 

It is worth to mention that, in Ref.\cite{Azreg-Ainou:2026xcc}, the authors also discussed chaotic imprints of dark matter in regular black hole background. However, their discussion focused on Simpson-Visser regular black holes.
As is mentioned in sec.\ref{sec1}, the regular black holes considered in this paper is constructed in conceptually different ways, leading to distinct physical interpretations of their parameters and of the role of dark matter. 
In this paper, regularity emerges dynamically from solving the Einstein equations with an anisotropic fluid source whose density profile follows a Dehnen-type dark matter halo. 
Importantly, the parameter $a$ has a clear physical meaning: it represents the characteristic scale of the dark matter halo and controls how the geometry transitions from the regular core to the outer vacuum region. 

Another key difference concerns the role of dark matter. In this paper, dark matter is the gravitational source that self-consistently determines the background metric via the field equations. In Ref.\cite{Azreg-Ainou:2026xcc}, the dark matter environment is modeled effectively as an external perturbative potential influencing particle dynamics, rather than being derived as the matter source of the spacetime geometry itself.

Next, we will discuss the motion of particle. To avoid the singularity of the metric in the event horizon, we would like to use the Painleve coordinate in the following discussion.
The line element in the Painleve coordinate is given by
\be\begin{aligned}\label{painlevemetric}
    ds^2=-f(r)dt^2+2\sqrt{1-f(r)}dt dr+dr^2+r^2 d\Omega^2
\end{aligned}\ee
with a coordinate transformation
\be\begin{aligned}
    dt\to dt-\frac{\sqrt{1-f(r)}}{f(r)}dr
\end{aligned}\ee

To study the influence of dark matter halos on the chaos of particle motion, we consider a massless test particle that is moving in an external harmonic potential.
For a general static, spherically symmetric metric of the form \eqref{painlevemetric}, the dispersion relation for a massless particle is the null constraint
\begin{equation}\label{eq:null_dispersion_general}
g^{\mu\nu}p_\mu p_\nu = 0,
\end{equation}
which is the $m\to 0$ limit of the normalization condition $p_\mu p^\mu=-m^2$ used for geodesic motion. The explicit form is given by
\begin{equation}
-p_t^2+2\sqrt{1-f(r)}\,p_r p_t+f(r)p_r^2+\frac{p_\theta^2}{r^2}+\frac{p_\phi^2}{r^2\sin^2\theta}=0.
\label{eq:null_dispersion_explicit}
\end{equation}
Static and spherical symmetry imply two conserved quantities along geodesics,
\begin{equation}
E\equiv -p_t,\qquad L\equiv p_\phi,\label{eq:conserved_EL}
\end{equation}
where $E$ is the energy measured at infinity and $L$ is the azimuthal angular momentum. Substituting \eqref{eq:conserved_EL} into \eqref{eq:null_dispersion_explicit} and solving for $E$ yields
\begin{equation}
E = \pm\sqrt{\frac{p_{\theta }^2+r^2 p_r^2+p_\phi^2 \csc ^2\theta}{r^2}}-\sqrt{1-f} p_r,
\label{eq:energy_massless_generaltheta}
\end{equation}
Noting that the energy has two branches. One represents ingoing particles, and the other represents outing ones. We will only consider the outgoing particle as is discussed in Ref.,i.e. choosing the $+$ sign.

For simplicity and clarity, we would like to study the particle with $L=0$ moving in a 2-dimensional external harmonic potential, which implies the total energy of this system can be written as 
\be\begin{aligned}
    E_{tot}=-\sqrt{1-f} p_r+\sqrt{\frac{p_{\theta }^2+r^2 p_r^2}{r^2}}+\frac{1}{2}K_r(r-r_0)^2+\frac{1}{2}K_\theta r_h^2(\theta-\theta_0)^2\,,
\end{aligned}\ee
where $r_0$ and $\theta_0$ are the center of the harmonic potential, $K_r$ and $K_\theta$ are the spring constant.
Then, we can easily obtain the Hamilton canonical equation of this system. In the regular black hole that we considered in this paper, the explicit form of the equation of motion is given by 
\be\begin{aligned}
    \dot r=&\frac{1}{2} r \left(\frac{2 p_r}{\sqrt{p_{\theta }^2+r^2 p_r^2}}-\left(\frac{a+2}{a+r}\right)^{3/2}\right)\,,\\
    \dot \theta=&\frac{p_{\theta }}{r \sqrt{p_{\theta }^2+r^2 p_r^2}}\,,\\
    \dot p_r=&\frac{(a+2)^{3/2} (2 a-r) p_r}{4 (a+r)^{5/2}}+\frac{p_{\theta }^2}{r^2 \sqrt{p_{\theta }^2+r^2 p_r^2}}+K_r \left(r_0-r\right)\,,\\
    \dot p_{\theta}=&K_\theta r_h^2  \left(\theta _0-\theta\right)
\end{aligned}\ee
where $\cdot$ denote the derivative with respect to $t$.
Given all the related parameter and initial conditions, we can solve this system numerically.

\begin{figure*}[t]
    \centering 
  
    \begin{subfigure}[b]{0.48\textwidth}
      \includegraphics[width=\linewidth]{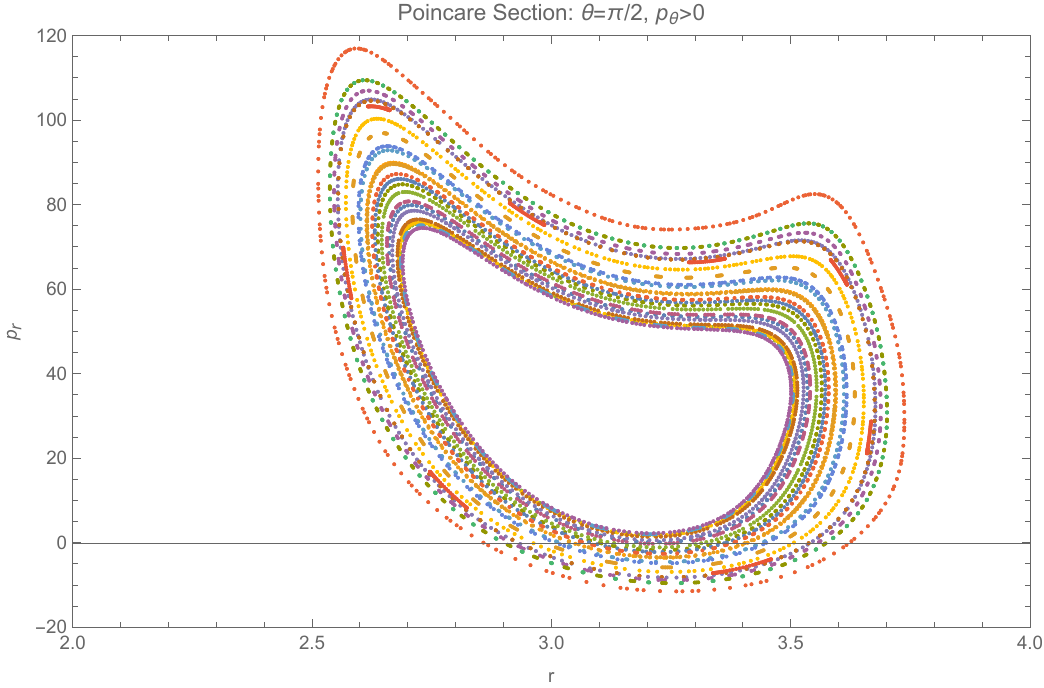}
      \caption{$a=0$}
      \label{fig:sub1}
    \end{subfigure}
    \begin{subfigure}[b]{0.48\textwidth}
      \includegraphics[width=\linewidth]{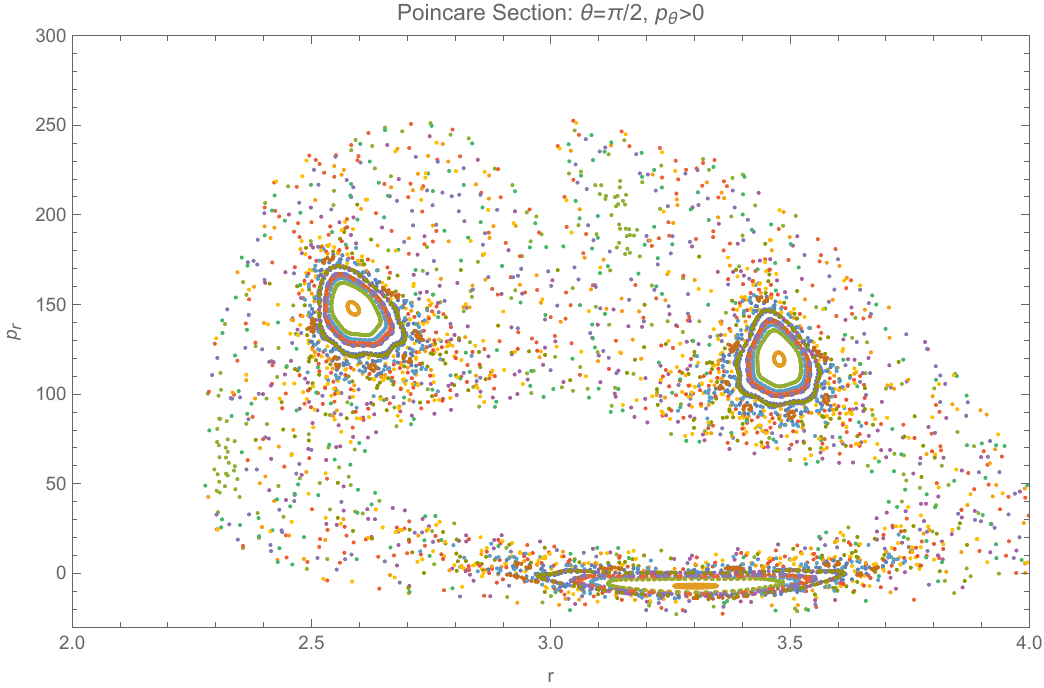}
      \caption{$a=1/8$}
      \label{fig:sub2}
    \end{subfigure}
    \\
    \begin{subfigure}[b]{0.48\textwidth}
      \includegraphics[width=\linewidth]{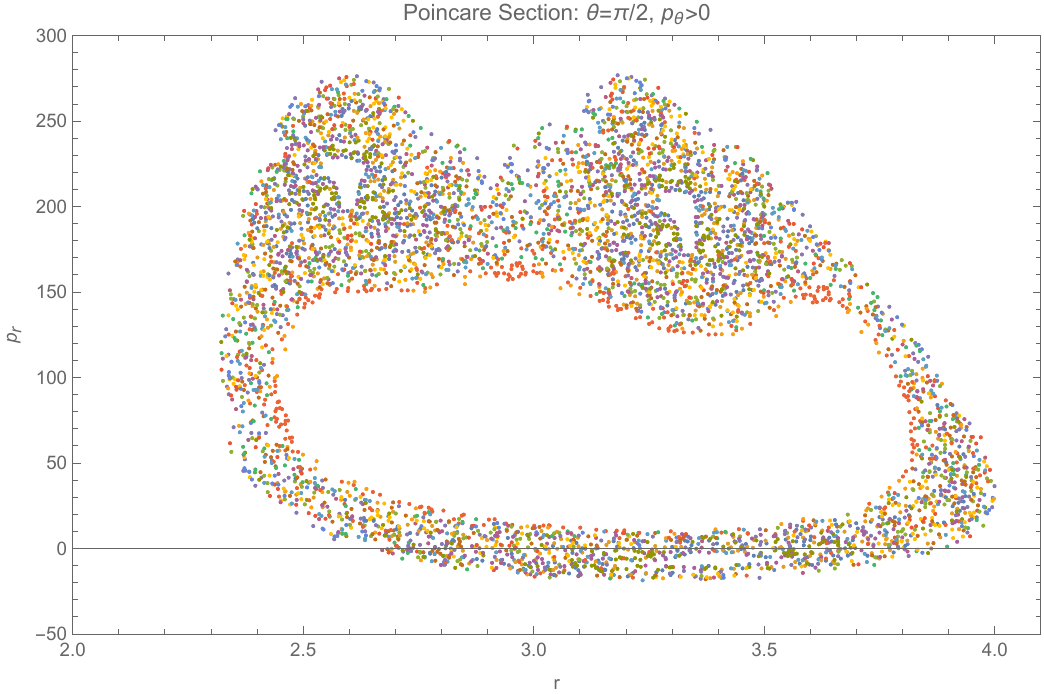}
      \caption{$a=1/4$}
      \label{fig:sub3}
    \end{subfigure}
    \begin{subfigure}[b]{0.48\textwidth}
      \includegraphics[width=\linewidth]{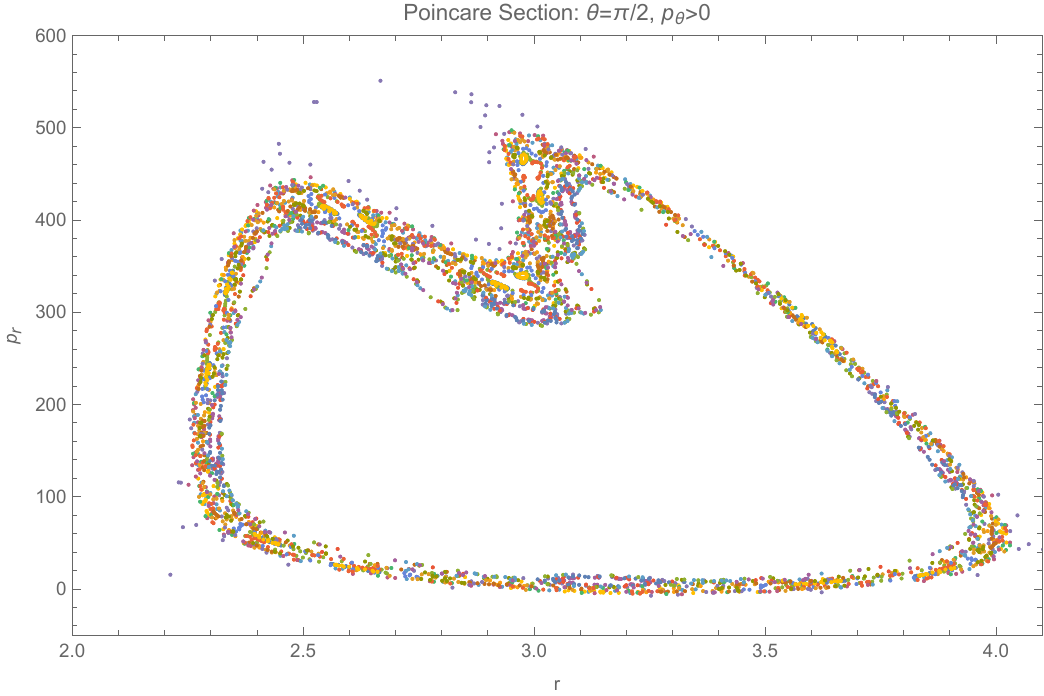}
      \caption{$a=1/2$}
      \label{fig:sub4}
    \end{subfigure}
  \\
      \begin{subfigure}[b]{0.48\textwidth}
      \includegraphics[width=\linewidth]{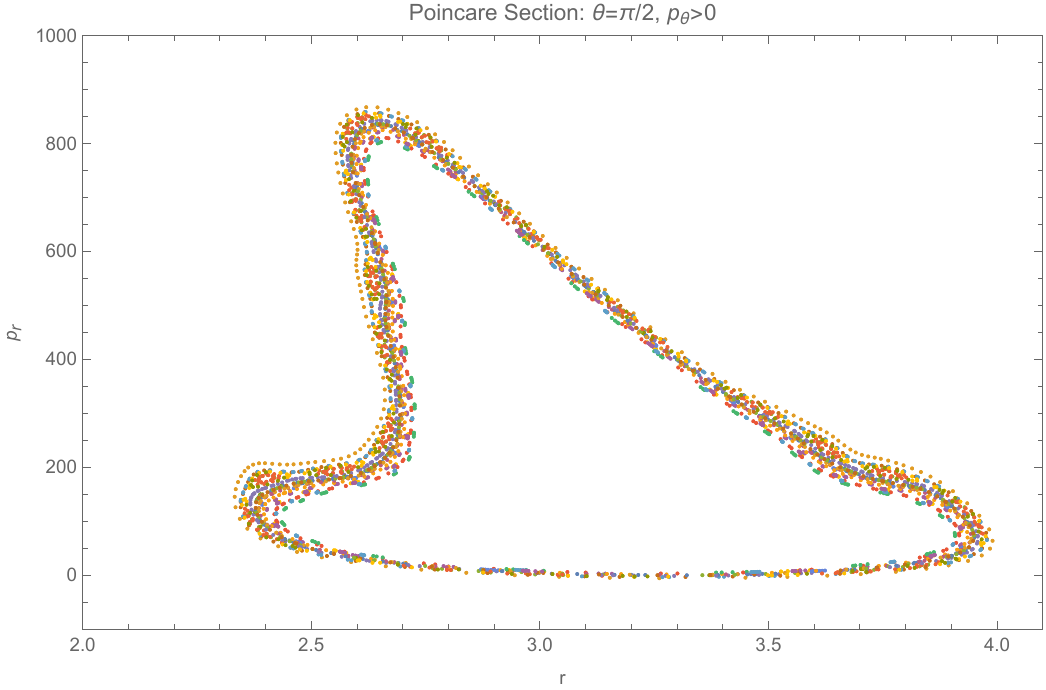}
      \caption{$a=3/4$}
      \label{fig:sub5}
    \end{subfigure}
        \begin{subfigure}[b]{0.48\textwidth}
      \includegraphics[width=\linewidth]{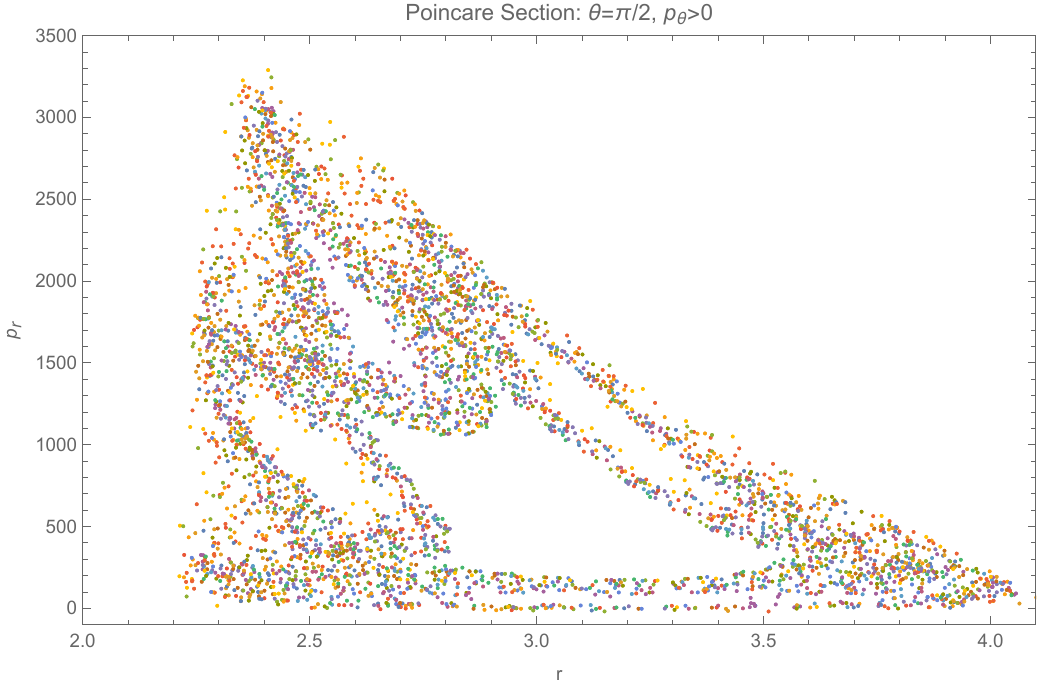}
      \caption{$a=1$}
      \label{fig:sub6}
    \end{subfigure}
    \caption{Poincar\'e sections recording particle crossings at $\theta=\pi/2$ with $p_\theta>0$. The energy is fixed to be $E=50$. The six panels correspond to different values of the halo scale parameter $a$, specifically $a=0, \frac{1}{8}, \frac{1}{4}, \frac{1}{2}, \frac{3}{4}, 1$. Initial conditions are randomly sampled within $r_{int} \in [3, 3.4]$, $p_r{}_{int} \in [-1, 1]$, and $\theta_{int} \in [13\pi/30, 17\pi/30]$, with distinct colors representing different initial condition. The harmonic potential center is located at $r_0=3.2$ and $\theta_0=\pi/2$ and the spring constant is $K_r=100$ and $K_\theta=25$.}
    \label{fig-1}
  \end{figure*}

\section{Poincar\'e section}\label{sec3}



In this section, we would like to study the chaotic motion in the spacetime of a regular black hole numerically. A powerful tool for studying chaotic phenomena is the Poincaré section, a geometric method that reduces the dimensionality of continuous dynamical systems by recording the sequence of intersections of phase space trajectories with a specific hypersurface, thereby transforming complex continuous motion into a set of discrete mapping points. In this paper, we choose the hypersurface to be $\theta=\pi/2$. Its key property lies in its ability to intuitively distinguish dynamical states: regular motion appears as orderly arranged points or closed curves on the section, while chaotic motion manifests as a cloud of irregularly distributed scattered points, thus clearly revealing the mechanism of the system's transition from order to chaos.

In Fig.\ref{fig-1}, we show the Poincaré sections for different values of the halo scale radius $a$.  We consider particles with energy $E=50$, ensuring that the changes observed in the Poincaré sections are solely due to variations in $a$, and not fluctuations in energy. To facilitate comparison with Ref.\cite{Dalui:2018qqv} , we character the external harmonic potential with $r_0=3.2$, $\theta_0=\pi/2$, $K_r=100$ and $K_\theta=25$. The initial condition are random points within $r_{int} \in [3, 3.4]$, $p_r{}_{int} \in [-1, 1]$, and $\theta_{int} \in [13\pi/30, 17\pi/30]$. The numerical calculations in this section were performed using \texttt{NDSolve} in \textit{Mathematica}. We employed the \texttt{ImplicitRungeKutta} method with \texttt{ImplicitRungeKuttaGaussCoefficients} to ensure energy conservation throughout the computation. And we set the maximum integration time as $t=5000$ in this section.

As shown in Fig.\ref{fig-1}, the Poincaré sections of the particle's motion change significantly with the increase of $a$. In Fig.\ref{fig:sub1}, corresponding to $a=0$, the motion is periodic, and the Poincaré section forms closed loops, characteristic of non-chaotic, regular orbits. This plot is consistent with the results of the Schwarzschild black hole with the same parameters, as shown in Ref.\cite{Dalui:2018qqv}. This behavior is consistent with the presence of KAM (Kolmogorov-Arnold-Moser) tori in the system. KAM tori represent stable, regular orbits in Hamiltonian systems and are a hallmark of non-chaotic motion. These tori are preserved in the system as long as the perturbation is small and the system remains in a nearly integrable state.

However, as $a$ increases, starting with a small $a$, see Fig.\ref{fig:sub2}, we observe that the previously periodic orbits begin to deform, and the Poincar\'e section starts to show more complex structures. This indicates the onset of chaos. The transition from regular motion to chaos is due to the increasing influence of the black hole's spacetime geometry, which is influenced by the dark matter halo described by the parameter $a$. As the parameter grows, the gravitational potential changes, leading to the breakdown of the KAM tori. The result is a departure from periodic motion to a more complex, aperiodic pattern.

In Figures\ref{fig:sub3}-\ref{fig:sub5}, corresponding to $a=1/4\,,1/2\,,3/4$, the Poincaré sections continue to show more intricate structures, indicating an increasing level of chaos. The periodic orbits, which are typically associated with the preservation of KAM tori, are progressively replaced by irregular and spread-out trajectories. This transition is a clear manifestation of deterministic chaos, where even small perturbations in the system lead to significantly different trajectories.

At $a=1$ (Fig.\ref{fig:sub6}), the Poincaré section displays a fully developed chaotic behavior, with no discernible closed loops or regular patterns. This chaotic regime is marked by trajectories that spread across the plot, signifying the complete destruction of the KAM tori and the emergence of sensitive dependence on initial conditions, a fundamental characteristic of chaos. As the KAM tori break apart, the system's dynamics become highly sensitive to small changes in initial conditions, leading to a state where predictability is lost.

The transition from periodic motion to chaotic dynamics is driven by the value of $a$, which controls the spacetime geometry and thus the gravitational potential influencing the particle's motion. The Poincaré sections captures the shift from regular orbits to chaotic trajectories as the parameter $a$ increases. These changes reflect the complex interactions between the particle's trajectory and the modified black hole geometry, which is itself influenced by the dark matter halo.







\section{Lyapunov Exponents}\label{sec4}

\begin{figure}[t]
    \centering
    \includegraphics[width=0.8\textwidth]{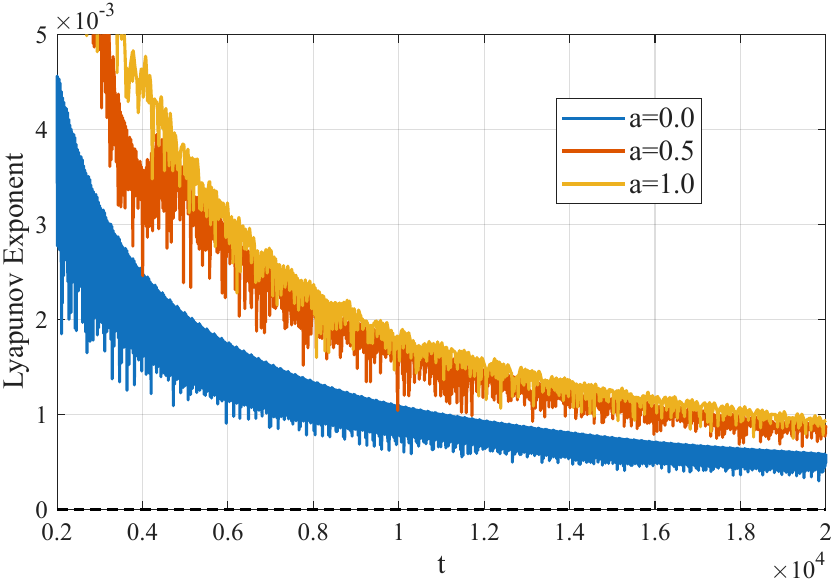}
    \caption{The Lyapunov exponent as a function of time for different values of the halo scale parameter $a$, specifically $a=0$(Blue), $a=1/2$(Red) and $a=1$(Orange).}\label{fig-lya}
  \end{figure}

In this section, we would like to investigate the Lyapunov exponents. 
In a chaotic system, even small differences between trajectories rapidly increase over time, making the long-term behavior of the system unpredictable.
The Lyapunov exponent is an important indicator used to describe the chaotic characteristics of dynamic systems, particularly in studying the system’s sensitivity to changes in initial conditions.
It measures the rate at which adjacent trajectories diverge as time progresses, thereby revealing whether the system exhibits chaotic behavior, defined by\cite{Dalui:2025bwm}
\be\begin{aligned}
    \lambda_{L,max}=\lim_{t\to\inf}\frac{1}{t}\ln\left(\frac{|\delta r(t)-r(t)|}{|\delta r(0)-r(0)|}\right)\,,
\end{aligned}\ee
where $\d r$ is the trajectory obtained after the initial conditions undergo a slight perturbation $|\delta r(0)-r(0)|$.
 In black hole dynamics, the Lyapunov exponent is often used to study the stability of particle motion, especially for particle orbits near black holes. A positive Lyapunov exponent typically indicates the presence of chaotic motion, which is crucial for understanding the dynamic properties around black holes and the potential observational effects (such as gravitational wave signals).

Fig.\ref{fig-lya} presents the Lyapunov exponent as a function of time for three different values of the halo scale parameter($a = 0, 0.5, 1$). As the parameter $a$ increases, the decay of the Lyapunov exponent slows down, which implies that higher values of $a$ leading to slower transitions to stability and a more persistent chaotic regime before settling down. Thus, the Lyapunov exponent captures the shift in the system's behavior from chaos to a more stable state as the halo parameter changes. Then, we calculated the corresponding maximum Lyapunov exponent and found $\lambda_{L,max}\approx 0.000505\,,\,0.000834 \,,\, 0.000902$ for $a=0\,,\,0.5\,,\,1$, which is consistent with the conclusions given by Fig.\ref{fig-1} and Fig.\ref{fig-lya}.


\section{Discussion}\label{sec5}

In this paper, we have studied the chaotic behavior of massless test particles near regular black holes supported by a Dehnen-type dark matter halo. We employed numerical techniques such as Poincaré sections and Lyapunov exponents to analyze the transition from regular to chaotic motion as the scale parameter of the dark matter halo increases. Our results show that dark matter significantly influences the dynamics of the particle orbits, with the halo acting as a primary driver of chaos in the system. We found that when the black hole is regular, the presence of dark matter leads to more chaotic particle motion. This finding is consistent with previous results, such as those in Ref.\cite{Das:2025eiv,Das:2025vja,Azreg-Ainou:2026xcc}, where the influence of dark matter on particle dynamics was also shown to enhance chaotic behavior near black holes. 
As the scale parameter $a$ increases, the transition from regular motion to chaotic dynamics becomes more evident, with the Lyapunov exponent indicating a growing instability in the system.



Our results provide valuable insights into the interactions between dark matter and black hole dynamics. They also offer a potential observational imprint for future studies, particularly in the context of gravitational wave signals. 
Future work could further explore the implications of these findings in observational settings

\acknowledgments
AS is supported by the NSFC (Grant No. 12250410250) and a Provincial Grant (Grant No. 2023QN10X389). AS would like to thank Jie Jiang and Qingxian Li for useful discussion.



\bibliographystyle{JHEP}
\bibliography{biblio.bib}

@article{Penrose:1964wq,
    author = "Penrose, Roger",
    title = "{Gravitational collapse and space-time singularities}",
    doi = "10.1103/PhysRevLett.14.57",
    journal = "Phys. Rev. Lett.",
    volume = "14",
    pages = "57--59",
    year = "1965"
}

@article{Hawking:1970zqf,
    author = "Hawking, S. W. and Penrose, R.",
    title = "{The Singularities of gravitational collapse and cosmology}",
    doi = "10.1098/rspa.1970.0021",
    journal = "Proc. Roy. Soc. Lond. A",
    volume = "314",
    pages = "529--548",
    year = "1970"
}

@book{Hawking:1973uf,
    author = "Hawking, Stephen W. and Ellis, George F. R.",
    title = "{The Large Scale Structure of Space-Time}",
    doi = "10.1017/9781009253161",
    isbn = "978-1-009-25316-1, 978-1-009-25315-4, 978-0-521-20016-5, 978-0-521-09906-6, 978-0-511-82630-6, 978-0-521-09906-6",
    publisher = "Cambridge University Press",
    series = "Cambridge Monographs on Mathematical Physics",
    month = "2",
    year = "2023"
}

@article{Bambi:2013caa,
    author = "Bambi, Cosimo and Malafarina, Daniele and Modesto, Leonardo",
    title = "{Non-singular quantum-inspired gravitational collapse}",
    eprint = "1305.4790",
    archivePrefix = "arXiv",
    primaryClass = "gr-qc",
    doi = "10.1103/PhysRevD.88.044009",
    journal = "Phys. Rev. D",
    volume = "88",
    pages = "044009",
    year = "2013"
}

@article{Hayward:2005gi,
    author = "Hayward, Sean A.",
    title = "{Formation and evaporation of regular black holes}",
    eprint = "gr-qc/0506126",
    archivePrefix = "arXiv",
    doi = "10.1103/PhysRevLett.96.031103",
    journal = "Phys. Rev. Lett.",
    volume = "96",
    pages = "031103",
    year = "2006"
}

@article{Ayon-Beato:1998hmi,
    author = "Ayon-Beato, Eloy and Garcia, Alberto",
    title = "{Regular black hole in general relativity coupled to nonlinear electrodynamics}",
    eprint = "gr-qc/9911046",
    archivePrefix = "arXiv",
    doi = "10.1103/PhysRevLett.80.5056",
    journal = "Phys. Rev. Lett.",
    volume = "80",
    pages = "5056--5059",
    year = "1998"
}

@article{Bronnikov:2000vy,
    author = "Bronnikov, Kirill A.",
    title = "{Regular magnetic black holes and monopoles from nonlinear electrodynamics}",
    eprint = "gr-qc/0006014",
    archivePrefix = "arXiv",
    doi = "10.1103/PhysRevD.63.044005",
    journal = "Phys. Rev. D",
    volume = "63",
    pages = "044005",
    year = "2001"
}

@article{Bronnikov:2005gm,
    author = "Bronnikov, K. A. and Fabris, J. C.",
    title = "{Regular phantom black holes}",
    eprint = "gr-qc/0511109",
    archivePrefix = "arXiv",
    doi = "10.1103/PhysRevLett.96.251101",
    journal = "Phys. Rev. Lett.",
    volume = "96",
    pages = "251101",
    year = "2006"
}

@article{Simpson:2018tsi,
    author = "Simpson, Alex and Visser, Matt",
    title = "{Black-bounce to traversable wormhole}",
    eprint = "1812.07114",
    archivePrefix = "arXiv",
    primaryClass = "gr-qc",
    doi = "10.1088/1475-7516/2019/02/042",
    journal = "JCAP",
    volume = "02",
    pages = "042",
    year = "2019"
}

@article{Fan:2016hvf,
    author = "Fan, Zhong-Ying and Wang, Xiaobao",
    title = "{Construction of Regular Black Holes in General Relativity}",
    eprint = "1610.02636",
    archivePrefix = "arXiv",
    primaryClass = "gr-qc",
    doi = "10.1103/PhysRevD.94.124027",
    journal = "Phys. Rev. D",
    volume = "94",
    number = "12",
    pages = "124027",
    year = "2016"
}

@article{Modesto:2008jz,
    author = "Modesto, Leonardo",
    title = "{Fractal Structure of Loop Quantum Gravity}",
    eprint = "0812.2214",
    archivePrefix = "arXiv",
    primaryClass = "gr-qc",
    doi = "10.1088/0264-9381/26/24/242002",
    journal = "Class. Quant. Grav.",
    volume = "26",
    pages = "242002",
    year = "2009"
}

@article{Ashtekar:2018lag,
    author = "Ashtekar, Abhay and Olmedo, Javier and Singh, Parampreet",
    title = "{Quantum Transfiguration of Kruskal Black Holes}",
    eprint = "1806.00648",
    archivePrefix = "arXiv",
    primaryClass = "gr-qc",
    doi = "10.1103/PhysRevLett.121.241301",
    journal = "Phys. Rev. Lett.",
    volume = "121",
    number = "24",
    pages = "241301",
    year = "2018"
}

@article{Bonanno:2000ep,
    author = "Bonanno, Alfio and Reuter, Martin",
    title = "{Renormalization group improved black hole space-times}",
    eprint = "hep-th/0002196",
    archivePrefix = "arXiv",
    reportNumber = "INFN-CT-03-00, MZ-TH-00-04",
    doi = "10.1103/PhysRevD.62.043008",
    journal = "Phys. Rev. D",
    volume = "62",
    pages = "043008",
    year = "2000"
}

@article{Navarro:1996gj,
    author = "Navarro, Julio F. and Frenk, Carlos S. and White, Simon D. M.",
    title = "{A Universal density profile from hierarchical clustering}",
    eprint = "astro-ph/9611107",
    archivePrefix = "arXiv",
    doi = "10.1086/304888",
    journal = "Astrophys. J.",
    volume = "490",
    pages = "493--508",
    year = "1997"
}

@article{Zhao:1995cp,
    author = "Zhao, HongSheng",
    title = "{Analytical models for galactic nuclei}",
    eprint = "astro-ph/9509122",
    archivePrefix = "arXiv",
    reportNumber = "MPA-885",
    doi = "10.1093/mnras/278.2.488",
    journal = "Mon. Not. Roy. Astron. Soc.",
    volume = "278",
    pages = "488--496",
    year = "1996"
}

@article{Konoplya:2025ect,
    author = "Konoplya, R. A. and Zhidenko, A.",
    title = "{Dark matter halo as a source of regular black-hole geometries}",
    eprint = "2511.03066",
    archivePrefix = "arXiv",
    primaryClass = "gr-qc",
    doi = "10.1103/7ptp-9j1t",
    journal = "Phys. Rev. D",
    volume = "113",
    number = "4",
    pages = "043011",
    year = "2026"
}

@article{Kar:2025phe,
    author = "Kar, Anjan and Kar, Sayan",
    title = "{Diverse regular spacetimes using a parametrised density profile}",
    eprint = "2504.12042",
    archivePrefix = "arXiv",
    primaryClass = "gr-qc",
    doi = "10.1140/epjc/s10052-025-14483-5",
    journal = "Eur. Phys. J. C",
    volume = "85",
    number = "7",
    pages = "773",
    year = "2025"
}

@article{Dekel:2017bwy,
    author = "Dekel, Avishai and Ishai, Guy and Dutton, Aaron A. and Maccio, Andrea V.",
    title = "{Dark-matter halo profiles of a general cusp/core with analytic velocity and potential}",
    eprint = "1610.00916",
    archivePrefix = "arXiv",
    primaryClass = "astro-ph.GA",
    doi = "10.1093/mnras/stx486",
    journal = "Mon. Not. Roy. Astron. Soc.",
    volume = "468",
    number = "1",
    pages = "1005--1022",
    year = "2017"
}

@article{Lutfuoglu:2026fks,
    author = {L{u}tf{u}oglu, Bekir Can},
    title = "{Particle Motion in Regular Black Hole Spacetimes Supported by a Galactic Halo}",
    eprint = "2602.16882",
    archivePrefix = "arXiv",
    primaryClass = "gr-qc",
    month = "2",
    year = "2026"
}

@article{Dalui:2018qqv,
    author = "Dalui, Surojit and Majhi, Bibhas Ranjan and Mishra, Pankaj",
    title = "{Presence of horizon makes particle motion chaotic}",
    eprint = "1803.06527",
    archivePrefix = "arXiv",
    primaryClass = "gr-qc",
    doi = "10.1016/j.physletb.2018.11.050",
    journal = "Phys. Lett. B",
    volume = "788",
    pages = "486--493",
    year = "2019"
}

@article{An:2025xmb,
    author = "An, Yu-Sen and Zhang, Wei-Hao",
    title = "{Probing quantum anomaly corrections on black hole physics through chaos}",
    eprint = "2512.19163",
    archivePrefix = "arXiv",
    primaryClass = "gr-qc",
    month = "12",
    year = "2025"
}

@article{Cao:2025qpz,
    author = "Cao, Deshui and Zhang, Lina and Chen, Songbai and Pan, Qiyuan and Jing, Jiliang",
    title = "{Chaotic motion of particles around a dyonic Kerr{\textendash}Newman black hole immersed in the Melvin-swirling universe}",
    eprint = "2511.08415",
    archivePrefix = "arXiv",
    primaryClass = "gr-qc",
    doi = "10.1140/epjc/s10052-025-15002-2",
    journal = "Eur. Phys. J. C",
    volume = "85",
    number = "11",
    pages = "1250",
    year = "2025"
}

@article{Dalui:2025bwm,
    author = "Dalui, Surojit and Bhattacharya, Soumya and Singha, Chiranjeeb",
    title = "{Probing chaos in Schwarzschild-de Sitter spacetime: The role of black hole and cosmological horizons}",
    eprint = "2502.12758",
    archivePrefix = "arXiv",
    primaryClass = "gr-qc",
    doi = "10.1016/j.physletb.2025.140028",
    journal = "Phys. Lett. B",
    volume = "872",
    pages = "140028",
    year = "2026"
}

@article{Cao:2024pdb,
    author = "Cao, Deshui and Zhang, Lina and Chen, Songbai and Pan, Qiyuan and Jing, Jiliang",
    title = "{Chaotic motion of particles in the spacetime of a Kerr black hole immersed in swirling universes}",
    eprint = "2410.03214",
    archivePrefix = "arXiv",
    primaryClass = "gr-qc",
    doi = "10.1140/epjc/s10052-025-13749-2",
    journal = "Eur. Phys. J. C",
    volume = "85",
    number = "1",
    pages = "28",
    year = "2025"
}

@article{He:2023dcz,
    author = "He, Yucheng and Wang, Zeqiang and Chen, Deyou",
    title = "{Report on chaos bound outside Taub-NUT black holes}",
    eprint = "2310.00315",
    archivePrefix = "arXiv",
    primaryClass = "gr-qc",
    doi = "10.1016/j.dark.2023.101325",
    journal = "Phys. Dark Univ.",
    volume = "42",
    pages = "101325",
    year = "2023"
}

@article{Chen:2016tmr,
    author = "Chen, Songbai and Wang, Mingzhi and Jing, Jiliang",
    title = "{Chaotic motion of particles in the accelerating and rotating black holes spacetime}",
    eprint = "1604.02785",
    archivePrefix = "arXiv",
    primaryClass = "gr-qc",
    doi = "10.1007/JHEP09(2016)082",
    journal = "JHEP",
    volume = "09",
    pages = "082",
    year = "2016"
}

@article{Azreg-Ainou:2026xcc,
    author = {Azreg-A{\"\i}nou, Mustapha and Jamil, Mubasher and Saridakis, Emmanuel N.},
    title = "{Chaotic imprints of dark matter in extreme mass-ratio inspirals}",
    eprint = "2602.19541",
    archivePrefix = "arXiv",
    primaryClass = "gr-qc",
    month = "2",
    year = "2026"
}

@article{Das:2025eiv,
    author = "Das, Surajit and Dalui, Surojit and Lee, Bum-Hoon and Cai, Yi-Fu",
    title = "{Extreme-Mass-Ratio Inspirals Embedded in Dark Matter Halo II: Chaotic Imprints in Gravitational Waves}",
    eprint = "2512.04848",
    archivePrefix = "arXiv",
    primaryClass = "gr-qc",
    month = "12",
    year = "2025"
}

@article{Das:2025vja,
    author = "Das, Surajit and Dalui, Surojit and Lee, Bum-Hoon and Cai, Yi-Fu",
    title = "{Extreme-Mass-Ratio Inspirals Embedded in Dark Matter Halo I:Existence of Homoclinic Orbit and Near-Horizon Chaos}",
    eprint = "2511.03657",
    archivePrefix = "arXiv",
    primaryClass = "gr-qc",
    month = "11",
    year = "2025"
}

@article{Taylor:2002zd,
    author = "Taylor, James E. and Silk, Joseph",
    title = "{The Clumpiness of cold dark matter: Implications for the annihilation signal}",
    eprint = "astro-ph/0207299",
    archivePrefix = "arXiv",
    doi = "10.1046/j.1365-8711.2003.06201.x",
    journal = "Mon. Not. Roy. Astron. Soc.",
    volume = "339",
    pages = "505",
    year = "2003"
}

@article{Dehnen:1993uh,
    author = "Dehnen, W.",
    title = "{A Family of Potential-Density Pairs for Spherical Galaxies and Bulges}",
    journal = "Mon. Not. Roy. Astron. Soc.",
    volume = "265",
    pages = "250",
    year = "1993"
}




\end{document}